\documentclass[superscriptaddress,showpacs,amsmath,amssymb,pra,twocolumn]{revtex4-1}

\usepackage{multibib}
\usepackage{tikz}
\usetikzlibrary{positioning}
\tikzset{>=stealth}
\usepackage{graphicx}
\usepackage{subcaption}

\usepackage{amsfonts}
\usepackage[english]{babel}
\usepackage{braket}

\usepackage[bookmarks=false,breaklinks,bookmarksopen,bookmarksnumbered,colorlinks,linkcolor=LINKCOL,linktocpage,citecolor=CITECOL,urlcolor=URLCOL,pdfpagemode=UseOutline,pdftex]{hyperref}

\usepackage{booktabs}
\usepackage{comment}

\definecolor{TITLECOL}{rgb}{0.1,0.2,0.7} 
\definecolor{SECOL}{rgb}{0.1,0.2,0.7} 
\definecolor{CONTENTSCOL}{rgb}{0.1,0.2,0.7} 
\definecolor{SSECOL}{rgb}{0.25,0,0.48} 
\definecolor{SSSECOL}{rgb}{0.2,0.08,0.53} 
\definecolor{FINCOL}{rgb}{0.01,0.3,0.07} 

\def\coloredtitle#1{\title{\textcolor{TITLECOL}{#1}}} 
\def\coloredauthor#1{\author{\textcolor{CITECOL}{#1}}} 

\definecolor{URLCOL}{rgb}{0,0.17,0.43} 
\definecolor{LINKCOL}{rgb}{0.05,0.4,0} 
\definecolor{CITECOL}{rgb}{0.35,0,0.48} 

\usepackage{natbib}
\def\bea{\begin{eqnarray}}
\def\eea{\end{eqnarray}}
\def\ben{\begin{equation}}
\def\een{\end{equation}}
\def\benu{\begin{enumerate}}
\def\enu{\end{enumerate}}

\def\bei{\begin{itemize}}
\def\eei{\end{itemize}}
\def\beit{\begin{itemize}}
\def\eit{\end{itemize}}
\def\benu{\begin{enumerate}}
\def\enu{\end{enumerate}}

\begin{document}

\coloredtitle{Theory for Nonlinear Spectroscopy of Vibrational Polaritons}

\coloredauthor{Raphael F. Ribeiro}
\affiliation{Department of Chemistry and Biochemistry, University of California San Diego, La Jolla, CA 92093}
\coloredauthor{Adam D. Dunkelberger}
\affiliation{Chemistry Division, Naval Research Laboratory, Washington, District Of Columbia 20375}
\coloredauthor{Bo Xiang}
\affiliation{Materials Science and Engineering Program, University of California, San Diego, La Jolla, CA 92093}
\coloredauthor{Wei Xiong}
\affiliation{Department of Chemistry and Biochemistry, University of California San Diego, La Jolla, CA 92093}
\affiliation{Materials Science and Engineering Program, University of California, San Diego, La Jolla, CA 92093}
\coloredauthor{Blake S. Simpkins}
\affiliation{Chemistry Division, Naval Research Laboratory, Washington, District Of Columbia 20375}
\coloredauthor{Jeffrey C. Owrutsky}
\affiliation{Chemistry Division, Naval Research Laboratory, Washington, District Of Columbia 20375}
\coloredauthor{Joel Yuen-Zhou}
\affiliation{Department of Chemistry and Biochemistry, University of California San Diego, La Jolla, CA 92093}
\date{\today}
\begin{abstract}
Molecular polaritons have gained considerable attention due to their potential to control nanoscale molecular processes by harnessing electromagnetic coherence. Although recent experiments with liquid-phase vibrational polaritons have shown great promise for exploiting these effects, significant challenges remain in interpreting their spectroscopic signatures. In this letter, we develop a quantum-mechanical theory of pump-probe spectroscopy for this class of polaritons based on the quantum Langevin equations and the input-output theory. Comparison with recent experimental data shows good agreement upon consideration of the various vibrational anharmonicities that modulate the signals. Finally, a simple and intuitive interpretation of the data based on an effective mode-coupling theory is provided. Our work provides a solid theoretical framework to elucidate nonlinear optical properties of molecular polaritons as well as to analyze further multidimensional spectroscopy experiments on these systems.
\end{abstract}
\maketitle

\def\am{a_{i}}
\def\amd{a^{\dagger}_{i}}
\def\ap{a_p}
\def\LP{_{\rm LP}}
\def\UP{_{\rm UP}}
\def\D{_{\rm D_\mu}}
\def\Dd{_{\rm D_\mu}^\dagger}
\def\Db{_{\rm D_{\overline{\mu}}}}
\def\hc{\text{h.c.}}\def\Oeff{\Omega_{\rm eff}}
\def\Deff{\Delta_{\rm{eff}}}
\def\ab{\alpha_{\rm{B}}}
\def\abd{\alpha_{\rm{B}}^\dagger}
\def\alp{\alpha\LP}
\def\aup{\alpha\UP}
\def\alpd{\alpha\LP^\dagger}
\def\aupd{\alpha\UP^\dagger}
\def\ext{_{\text{ext}}}
\def\pr{^{\text{pr}}}
\def\pu{^{\text{pu}}}
\def\nonum{\nonumber \\}
\def\in{\text{in}}
\def\out{\text{out}}
\def\cm{~\text{cm}^{-1}}

\emph{Introduction.\textendash{}} The strong coupling regime between optical microcavities and molecular vibrational modes has been recently achieved with polymers \cite{shalabney_coherent_2015, long_coherent_2015, simpkins_spanning_2015, ahn_vibrational_2017}, proteins \cite{vergauwe_quantum_2016}, liquid-phase solutions \cite{thomas_ground-state_2016, casey_vibrational_2016, dunkelberger_modified_2016} and neat liquids \cite{george_liquid-phase_2015}. These works have unambiguously shown the existence of a novel class of hybrid excitations (vibrational polaritons) consisting of superpositions of delocalized molecular vibrations and microcavity electromagnetic (EM) field modes \cite{ebbesen_hybrid_2016}. It is now clear that the dynamics of vibrational polaritons is substantially different than that of bare molecular excited-states \cite{dunkelberger_modified_2016, chervy_vibro-polaritonic_2017}. Given that polariton properties are tunable, the strong coupling of light and vibrational degrees of freedom opens up new routes for the control of chemical processes \cite{shapiro2003principles,agranovich_hybrid_2011, sanvitto_road_2016, ebbesen_hybrid_2016}.
 \begin{figure}[b]
\includegraphics[scale=0.4]{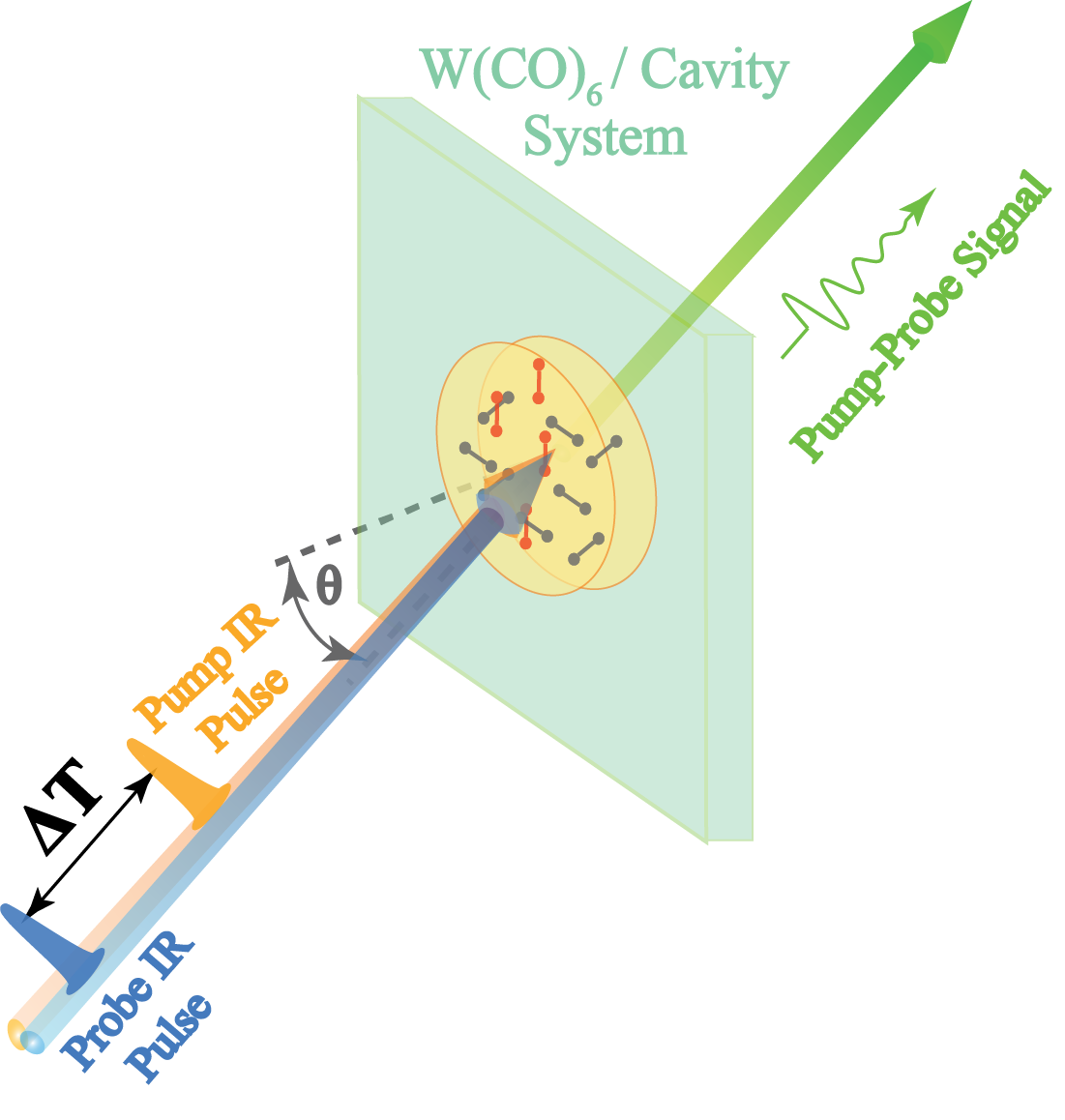}
\caption{Experimental setup for pump-probe spectroscopy of polaritons formed from W(CO)$_6$ molecules embedded in an optical microcavity} \label{setup}
\end{figure}
Vibrational polaritons can be detected whenever a microcavity mode and a molecular vibration exchange energy at a rate that is faster than their dephasing rates \cite{ebbesen_hybrid_2016}. Notably, the liquid-phase vibrational state contributing to the formation of polaritons is effectively dispersionless, and under weak coupling, essentially localized. Thus, a significant feature of the strong coupling regime is the introduction of a mesoscopic coherence length on the material infrared (IR) polarization \cite{aberra_guebrou_coherent_2012,agranovich_hybrid_2011,spano_optical_2015}. This key point makes polaritons substantially different to conventional molecular states. Similar is true in the case of disordered organic excitons in molecular aggregates: their excitations are at most delocalized about not more than a few hundreds of chromophores \cite{knoester_modeling_2006,huh_atomistic_2014} due to dipolar interactions but, upon strong coupling a coherence length on the order of $\mu m$ is observed \cite{aberra_guebrou_coherent_2012,agranovich_hybrid_2011}. In fact, the field of organic semiconductor cavity polaritons has seen a surge in the last 15 years, with a variety of exciting phenomena demonstrated experimentally \cite{virgili_ultrafast_2011, aberra_guebrou_coherent_2012, hutchison_tuning_2013, daskalakis_nonlinear_2014, coles_polariton-mediated_2014, orgiu_conductivity_2015, lerario_room-temperature_2017, zhong_energy_2017}, and interesting theoretical predictions awaiting observation \cite{feist_extraordinary_2015, schachenmayer_cavity-enhanced_2015, yuen-zhou_plexciton_2016, herrera_cavity-controlled_2016, bennett_novel_2016, saurabh_two-dimensional_2016, cortese_collective_2017, flick_cavity_2017, martinez-martinez_can_2017}.
 However, despite the similarity in the linear response of electronic and vibrational polaritons, there exists obvious differences in their nonlinear dynamics. In particular, vibrational modes can be well-approximated at low energies by weakly-perturbed harmonic oscillators, while the same is generically not true for the electronic degrees of freedom. Thus, intriguing phenomena have been recently observed under liquid-phase vibrational strong-coupling without counterpart in exciton-polariton systems. This includes e.g, the suppression and change of mechanism of an electronic ground-state chemical reaction \cite{thomas_ground-state_2016}, surprisingly enhanced Raman scattering \cite{shalabney_enhanced_2015,del_pino_signatures_2015,strashko_raman_2016} and, among other features, the significant derivative lineshape and wave-vector dependence of vibrational polariton dynamics in the cavity-W(CO)$_6$ pump-probe (PP) spectra reported by Dunkelberger et al \cite{dunkelberger_modified_2016}. Nonlinear spectroscopy is a particularly important tool to uncover the fundamental properties of vibrational polaritons \cite{mukamel1999principles} as it directly probes excited-state dynamics and thus the vibrational anharmonicity without which chemical reactions would not occur. Motivated by the intriguing PP spectra of vibrational polaritons first reported in ref. \cite{dunkelberger_modified_2016}, and the recent 2D spectra of Xiang et al. \cite{bo_xiang_revealing_2017}, we present in this letter a quantum-mechanical model for vibrational polariton PP spectroscopy including the effects of both mechanical and electrical anharmonicity.

\emph{Theory of pump-probe response of vibrational polaritons. \textendash{}} We consider a setup with $N$ identical independent vibrational (molecular) degrees of freedom strongly-coupled to a single-mode planar [Fabry-Perot (FP)] microcavity \cite{kavokin2017microcavities} (Fig. 1). The former are assumed to be weakly-coupled to intramolecular modes and solvent degrees of freedom (bath) while the latter is weakly-coupled to the external (vacuum) EM modes on the left- and right-hand sides of its transverse direction. Thus, the total Hamiltonian is given by a sum of two contributions $H = H_0 + H_{\text{anh}}$: $H_0$ is the zeroth-order harmonic Hamiltonian
\begin{align}  H_0 = \omega_0 \sum_{i=1}^N \amd \am +\omega_c b^\dagger b - g_1  \sum_{i=1}^{N}\left(\amd b + b^\dagger \am\right) + H_{\text{SB}}, \end{align}
where we choose units such that $\hbar=1$, $\omega_0 (\omega_c)$ is the molecular (cavity mode) fundamental frequency, $\am$ ($b$) is the $i$th molecule (cavity mode) annihilation operator, $g_1$ is the (real) single-molecule light-matter coupling and $H_{\text{SB}}$ is the system-bath interaction which contains the linear coupling between cavity and external modes, and the interaction between the molecular vibrations and their (independent) baths (see SI, eq. 1). In the response formalism detailed below the effects of $H_{\text{SB}}$ on the molecular system are accounted for with a phenomenological damping constant $\gamma_m$ which can be determined by fitting the experimentally-available bare vibrational fundamental transition lineshape to a Lorentzian distribution. The anharmonic part of the total Hamiltonian is given by:

\begin{align} H_{\text{anh}} = -\Delta \sum_{i=1}^{N}\amd \amd \am \am - g_3 \sum_{i=1}^N \left(b^\dagger \amd \am \am + \amd \amd \am b\right), \label{han} \end{align}
where $\Delta > 0$ implies the $1\rightarrow2$ vibrational transition is red-detuned from $\omega_0$ by $2\Delta$, while $g_3$ can be either positive or negative as it represents the deviation of the vibrational $1\rightarrow 2$ transition dipole moment ($\mu_{1\rightarrow 2}$) from the corresponding result for the harmonic oscillator. In particular, $g_3$ manifests itself by the relation $\mu_{1\rightarrow2} = \sqrt{2}\mu_{0\rightarrow1}(1+g_3/g_1)$ \cite{khalil_coherent_2003,khalil_signatures_2001} so that if $g_3$ and $g_1$ have the same (opposite) sign, the $\mu_{1\rightarrow 2}$ of the anharmonic system will be larger (smaller) relative to that of a harmonic. Because $\Delta$ parametrizes the anharmonicity of molecular motion, we denote it by \textit{mechanical (nuclear) anharmonicity}. The other term in Eq. \ref{han} is called \textit{electrical anharmonicity} \cite{herzberg1939molecular, mccoy_vibrational_2012}, for it may be understood to represent a deviation from harmonic behavior in the interaction between molecular vibrations and the electric field of light.

We employ input-output theory \cite{gardiner_input_1985, portolan_nonequilibrium_2008, carusotto_quantum_2013, li_probing_2017} to estimate the PP transmission spectrum of vibrational polaritons. The strategy of the input-output method is to relate the microcavity EM field modes at a time $t$ with the state of the external EM field at an earlier and later times $t_0 < t$ and $t_1 > t$, respectively \cite{gardiner_input_1985, portolan_nonequilibrium_2008,carusotto_quantum_2013}. The earlier and later external EM field modes are denoted input and output fields, respectively. For instance, the Langevin-Heisenberg equation of motion for the cavity mode annihilation operator (in the appropriate rotating-wave approximation) is given by
\begin{equation} \frac{\mathrm{d}}{\mathrm{d}t} b(t) = -i\omega_c b(t) -\frac{\kappa}{2}b(t)  - \sqrt{\frac{\kappa}{2}}b_{\in}^{\text{L}}(t) - i g_1 P_1(t) - ig_3 P_3(t), \label{bt} \end{equation}
where $\kappa/2$ is the cavity linewidth due to the coupling between external and cavity EM modes, and $b_{\in}^{\text{L}}(t)$ is the l.h.s cavity input operator (we assume the input laser pulse interacts with the cavity from the left). Similarly, the molecular polarization operators $P_1(t) =\sum_i a_i(t)$  and $P_3(t) = \sum_i a_i(t)^\dagger a_i(t) a_i(t)$ (see Eqs. 1 and 2)  satisfy
\begin{align} & \frac{\mathrm{d}P_1(t)}{\mathrm{d}t} = -i\omega_0 P_1(t) -\frac{\gamma_m}{2}P_1(t)+ 2i \Delta P_3(t) - i g_1Nb(t)  \nonum
& -2ig_3 \sum_{i=1}^N (\amd\am b)(t) - ig_3b^\dagger(t)\sum_{i=1}^N(\am\am)(t), \label{p1t} \\
& \frac{\mathrm{d}P_3(t)}{\mathrm{d}t}=-i(\omega_0-2\Delta)P_3(t) -\frac{3\gamma_m}{2} P_3(t) \nonum
& -2i(g_1+g_3)\sum_{i=1}^N (\amd\am b)(t) -3ig_3\sum_{i=1}^N (\amd\amd\am\am b)(t)\nonum
 &+i g_1 b^\dagger(t) \sum_{i=1}^N (\am \am)(t)- 2\Delta\sum_{i=1}^N (\amd\amd\am\am\am)(t) \label{p3t},
\end{align}
where $\gamma_m > 0$ is the full width at half-maximum of the molecular vibrations induced by interaction with a local bath. Eqs. 3-5 are obtained by integrating out the external EM field and molecular bath degrees of freedom on the reasonable assumptions of weak system-bath couplings and Markovian dynamics \cite{gardiner_input_1985, gardiner1991quantum, nitzan2006chemical,steck2007quantum}. Note that the input and output operators satisfy the following useful relation \cite{gardiner_input_1985, gardiner1991quantum, steck2007quantum}
\begin{equation} b_{\out}^{\text{L}}(t) - b_{\in}^{\text{L}}(t) = \sqrt{\frac{\kappa}{2}} b(t) =  b_{\out}^{\text{R}}(t) - b_{\in}^{\text{R}}(t) \label{inout},\end{equation}
which allows us to obtain the output EM field given the microcavity response and the input field.
Given that transmission and reflection spectra can be obtained as Fourier transforms of the autocorrelation function of the external electric fields \cite{steck2007quantum} the approach we utilize here directly relates the observed spectra  to the dynamics of the system.  

In this letter, we focus on the pump-induced probe transmission observed at sufficiently long waiting (PP time delay) time $T$ \cite{mukamel1999principles,yuen2014ultrafast}, such that coherences have dephased and thus only transient molecular population variables need to be retained for an accurate description of the optical response. In other words, we assume the only significant effect of the pump on the system at the probe delay time is the generation of a transient molecular excited-state population inside the cavity. We also neglect any photonic population at the PP time delay. This assumption is consistent with the fact that in the experiments which we ultimately compare our theory to the cavity photon lifetime and probe delay are approximately 5 and 25 ps, respectively \cite{bo_xiang_revealing_2017, dunkelberger_modified_2016}. Thus, the last two terms of Eq. \ref{p3t} can be neglected, as their averages tend to zero at the PP time delay $T$. We also assume that $\braket{\amd\amd\am\am}(t)$ is much smaller than $\braket{\amd\am}(t)$ for $t \approx T$. This is reasonable, as we expect that after pumping the system with a classical field, each factor of $\am(t)$ contributes a factor of $\approx e^{-\gamma_m t/2}$ to these expectation values. In view of the listed conditions we only need a single parameter (which is assumed to be constant for the waiting times of interest)
\begin{equation} f\pu = \frac{\sum_{i=1}^{N} \braket{\amd \am}(T)}{N}, \end{equation}
to represent the effects of the pump on the system at the time where the probe acts. In this case, $\braket{P_3}(t)$ can be quickly obtained as a function of $f\pu$,
\small
\begin{align}  \frac{\mathrm{d}\braket{P_3}(t)}{\mathrm{d}t} =& -i(\omega_0-2\Delta)\braket{P_3}(t) -2i(g_1+g_3)Nf\pu \braket{b}(t),\\
&\braket{P_3}(\omega') =  N f\pu A_3(\omega') \braket{b}(\omega'), \nonumber \\
&A_3(\omega') = \frac{2(g_1+g_3)}{\omega'-(\omega_0-2\Delta)+i\gamma_3}, \end{align}
\normalsize
where $\braket{B}(t)$ is the expectation value of the Heisenberg operator $B(t)$, $\gamma_3 = 3\gamma_m/2$ and $\braket{B}(\omega')$ is the expectation value of the Fourier transform of $B(t)$.
Similarly, we obtain the following approximation for $\braket{P_1}(\omega')$,
\begin{align} &\braket{P_1}(\omega') = A_1\left(\omega',f\pu\right) N \braket{b}(\omega'), \\ & A_1\left(\omega',f\pu\right) = \frac{g_1+2g_3f\pu-2\Delta f\pu A_3(\omega')}{\omega' -\omega_0 +i\gamma_1},
\end{align}
where $\gamma_1 = \gamma_m/2$. Note that if $f\pu = 0$, then $A_1 (\omega',0) = g_1/(\omega'-\omega_0 + i\gamma_1)$, which is the appropriate result in the absence of a pump-induced excited-state population. From Eq. \ref{inout} we can now obtain an approximation to the absolute pump-probe linear transmission spectrum $ T^{\text{pu}}(\omega') = |\braket{b_{\out}^\text{R}}(\omega')|^2$ in terms of the transient matter polarization components $\braket{P_1}(\omega')$ and $\braket{P_3}(\omega')$, and the input probe field $\braket{b_\in^L}(\omega')$, via
\begin{widetext}
\begin{align} & \braket{b_{\out}^{\text{R}}}(\omega') = \frac{-i\frac{\kappa}{2} (\omega'-\omega_0 +i\gamma_1)(\omega'-(\omega_0-2\Delta)+i\gamma_3)}{(\omega'-\omega_c +i \kappa/2)(\omega'-\omega_0 +i\gamma_1)(\omega'-(\omega_0-2\Delta)+i\gamma_3)-N A(\omega',f\pu)}\braket{b_{\text{in}}^L}(\omega'),\label{ppfinal}
\end{align}
\end{widetext}
where
\begin{align} & A(\omega',f\pu) = g_1^2[\omega'-(\omega_0-2\Delta)+3i\gamma_m/2] + f\pu B(\omega'),  \nonumber \end{align}
\begin{align}
B(\omega') = & 4g_1g_3\left[\omega'-\omega_0 + i\gamma_m\right] +2g_3^2(\omega'-\omega_0+i\gamma_m/2)  \nonum
 &- 4\Delta g_1^2.\end{align}
Eq. \ref{ppfinal} is the central result of this work. It expresses the transmission spectrum of a cavity strongly-coupled with vibrational modes, given a pump-induced vibrational excited-state population $(Nf\pu)$. Note that if $g_3 = \Delta = 0$ no third-order signal exists, in consistency with the fact that harmonic systems exhibit no nonlinear response \cite{mukamel_quantum_2011}.

\begin{figure}[h]
\includegraphics[scale=0.5]{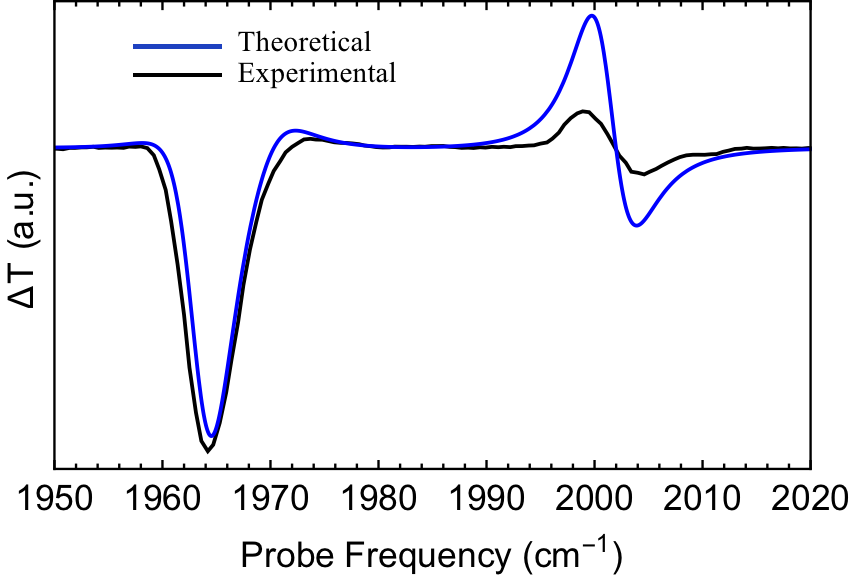}
\caption{25 ps experimental (black) \cite{bo_xiang_revealing_2017} and theoretical (blue) PP spectrum obtained assuming $f\pu=0.075$ and $\omega_c  = \omega_0 = 1983 \cm$.} \label{expth}
\end{figure}

\emph{Comparison to experimental results.\textendash{}}
We compare the predictions of our model with experiments which utilized solutions of W(CO)$_6$ in hexane \cite{dunkelberger_modified_2016, bo_xiang_revealing_2017}. The triply-degenerate carbonyl asymmetric stretch $T_{1u}$ of the referred molecule was chosen to be strongly-coupled to a resonant IR FP cavity. For this system, $ \omega_c = \omega_0 = 1983\cm$, $\gamma_m = 3\cm$, $\kappa = 11\cm$, $\Delta  = 7.5\cm$ and $g_1\sqrt{N} = 19\cm$. The electrical anharmonicity parameter is unknown for this molecule. Thus, we choose $g_3/g_1 = -0.25$, as a similar value was reported by Khalil et al. \cite{khalil_coherent_2003} for the carbonyl stretch in a different system. Additionally, we take $f\pu=0.075$ in all reported results (for a discussion of polariton and dark-state absorption, see SI). In Fig. \ref{expth} experimental and theoretical results are compared for the polariton PP (differential transmission) spectrum $\Delta T(\omega)$,
\begin{equation}
\Delta T(\omega) = T\pu(\omega) - T^0(\omega),
\end{equation}
where  $T\pu(\omega)$ is the transmission spectrum after excitation of the system with the pump, and  $T^0(\omega)$ is the linear transmission obtained in the absence of pumping. The experimental spectrum \cite{bo_xiang_revealing_2017} was obtained after a probe delay time of 25 ps (see also \cite{dunkelberger_modified_2016}). We recall that for the theoretical model $T\pu(\omega) = |\braket{b^\text{R}_{\text{out}}}(\omega)|^2|_{f\pu=0.075}$,~while $T^0(\omega)=|\braket{b^\text{R}_{\text{out}}}(\omega)|^2|_{f\pu=0}$. 

\begin{figure}[t]
\centering
\includegraphics[width=\columnwidth]{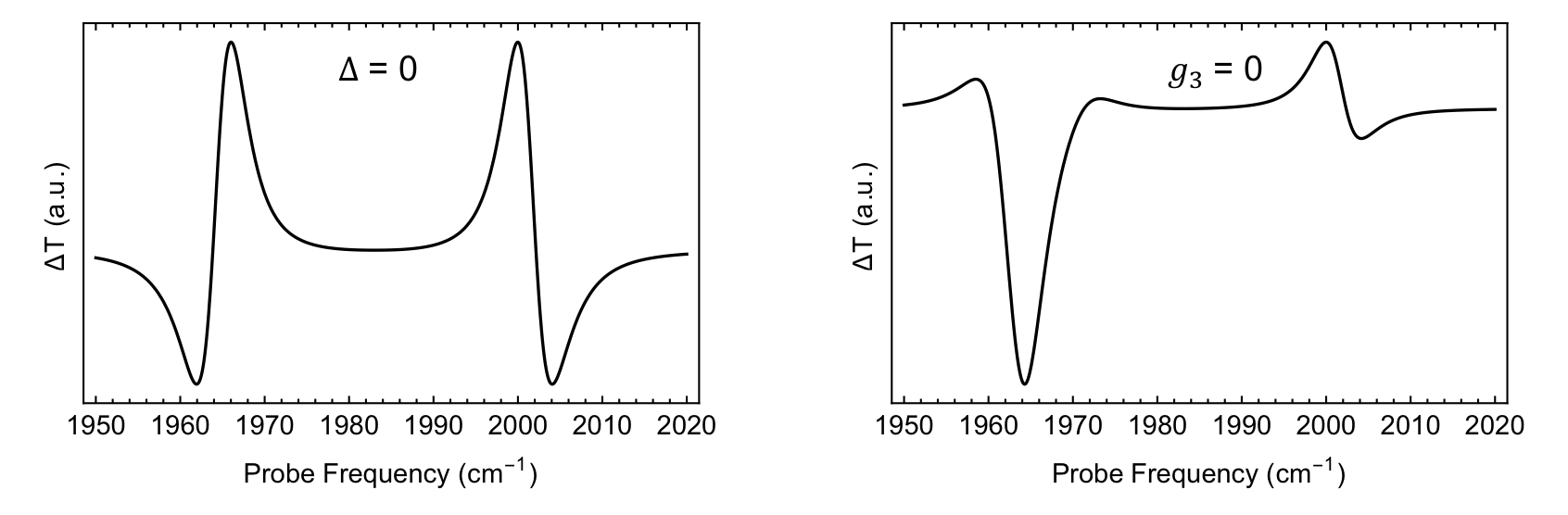}
\caption{Theoretical PP spectrum with vanishing mechanical (l.h.s) or electrical anharmonicity (r.h.s).}\label{elnucan}
\end{figure}

The two dominant features of the experimental PP spectrum are reproduced by our theory: the large negative feature in a neighborhood of the linear LP frequency, and the red-shift of the UP resonance. The theoretical prediction for the intensity of the latter is overestimated compared to the experimental result. This can be attributed to the various approximations employed in the derivation of Eq. \ref{ppfinal}. To understand the effects of vibrational anharmonicity on the polariton spectrum, we show in Fig. \ref{elnucan} the theoretical PP spectrum obtained when either electrical (l.h.s) or mechanical anharmonicity (r.h.s) are turned off.
Electrical anharmonicity gives rise to a transient spectrum with blue(red)-shift for the LP(UP) in conformity with the notion that it reduces the effective light-matter interaction (compared to a harmonic model). However, for systems with $\mu_{1\rightarrow 2} > \sqrt{2}\mu_{0\rightarrow 1}$ the Rabi splitting would be increased. Notably, the transmission peaks of LP and UP remain symmetrically distributed around the fundamental frequency. Mechanical anharmonicity also has simple effects on the PP spectrum: the UP red-shifts, and the LP resonance splits into two (l.h.s of Fig. 2). This is explained by the fact that  $\Delta > 0$ gives rise to red-detuned overtone frequencies relative to the fundamental. Note that
while the theoretical spectrum with $g_3 = 0$ resembles the experimental in Fig. 1, the former contains a positive bump near 1960$\cm$ which is absent from the latter. Thus the best agreement with experiment is observed when both electrical and nuclear anharmonicities are included in the theoretical model.

\begin{figure}[t]
\centering
\includegraphics[width=\columnwidth]{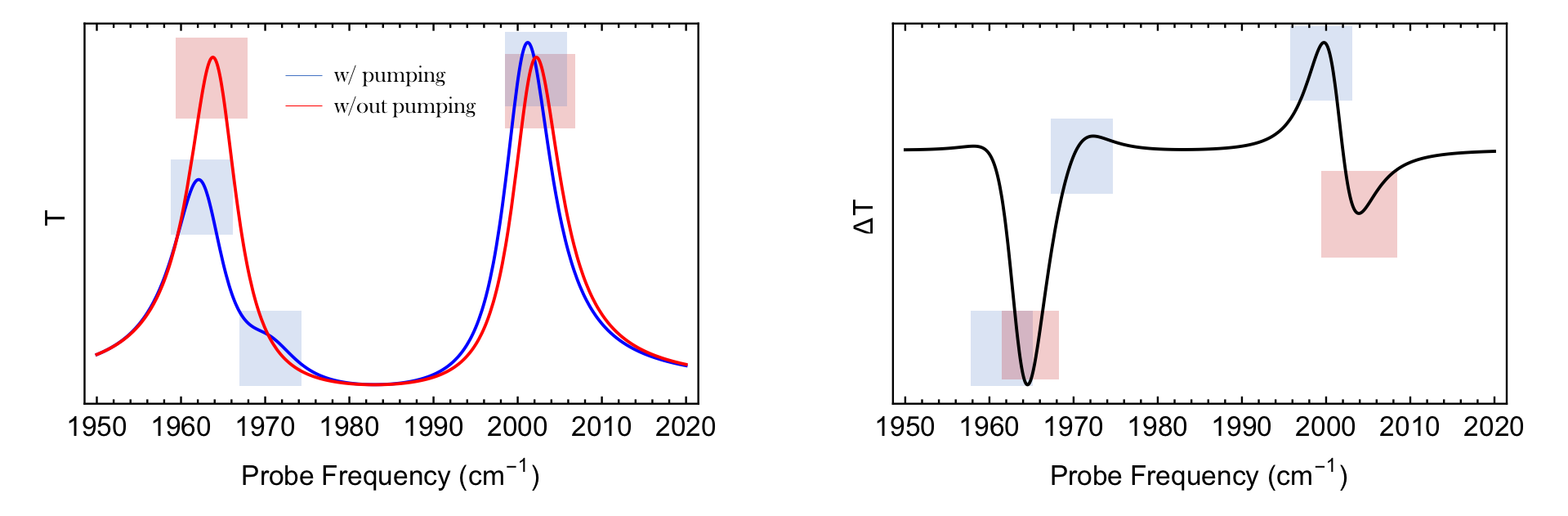}
\caption{l.h.s: Theoretical transmission spectrum in the presence (blue) and absence of pumping (red); r.h.s: Theoretical PP spectrum (difference between blue and red curves of l.h.s) with identified resonances} \label{theo-in}
\end{figure}

At this point we note that when $\gamma_m \rightarrow 0$ (for the case with significant molecular damping see SI), the resonances of the PP transmission spectrum  (Eq. \ref{ppfinal}) can be obtained as eigenvalues of the mode-coupling matrix
\footnotesize
\begin{equation}
h(f\pu)=  \begin{pmatrix}
\omega_c-i\kappa/2 & g_1 \sqrt{N}\sqrt{(1-2f\pu)} & g_2\sqrt{2f\pu N} \\
g_1 \sqrt{N}\sqrt{1-2f\pu} & \omega_{01} & 0 \\
g_2\sqrt{2f\pu N} & 0 & \omega_{12}
\label{mcmatrix}
 \end{pmatrix},
\end{equation}
\normalsize
 where $g_2 = g_1 + g_3$. The diagonal elements correspond to the bare cavity photon and matter excitation frequencies $ \omega_{01} = \omega_0$ and $\omega_{12} = \omega_0 -2\Delta$. Thus, we may assign the label 1 to the cavity photon, and 2 and 3 to the matter polarization components $P_{0\leftrightarrow1}$ and  $P_{1\rightarrow 2}$. $P_{0\leftrightarrow1}$ represents the effective material polarization due to stimulated emission and ground-state bleach, and $P_{1\rightarrow 2}$ is the excited-state absorption contribution. Hence, the off-diagonal elements of $h(f\pu)$ can be interpreted as couplings between the cavity and the different components of matter polarization. Note the interaction between $P_{1\rightarrow 2}$ and the cavity photon depends linearly on $\sqrt{f\pu N}$, while that between $P_{0\leftrightarrow 1}$ and the cavity depends on $\sqrt{(1-f\pu)N - f\pu N}$. The reason for this unusual expression for the coupling between $P_{0\leftrightarrow 1}$ and light is that the former may be viewed to arise from interference of polarizations due to stimulated emission (which depends on $f\pu N$), and ground-state bleach, which depends on $(1-f\pu)N$ (for additional details, including the corresponding double-sided Feynman diagrams, see Sl). When $f\pu > 1/2$, stimulated emission provides the dominant contribution to $P_{0\leftrightarrow 1}$, and this leads to non-hermiticity in Eq. \ref{mcmatrix} even when $\kappa = 0$.  Nonetheless, the complex eigenvalues of  $h$ are equal to the poles of the response function in Eq. \ref{ppfinal} even when $f\pu > 1/2$. Ultimately, Eq. \ref{mcmatrix} gives a simple interpretation of the three resonances appearing in the vibrational polariton transmission spectrum in the presence of a pump-induced incoherent excited-state population (Fig. \ref{theo-in}): the PP resonances correspond to \textit{transient polaritons} formed by the linear coupling of cavity photons with the matter polarization associated to transitions from the molecular ground- and excited-states. 

Notably, the mode-coupling matrix in Eq. \ref{mcmatrix} could have been obtained within a purely classical model of the optical response by assigning the $P_{0\leftrightarrow1}$, and $P_{1\rightarrow2}$ parts of the matter polarization to  the totally symmetric mode of $(1-f\pu)N$ and $Nf\pu$ Lorentz (harmonic) oscillators with frequencies $\omega_0$ and $\omega-2\Delta$, and single-mode electromagnetic field couplings $g_1$, and $g_3$, respectively. Thus, the assumptions we took to obtain Eq. \ref{ppfinal} provide the connection of the quantum-mechanical polariton theory with the classical model employed in the interpretation of vibrational polariton pump-probe spectra in Ref. \cite{dunkelberger_modified_2016}. The QM picture resulting from Eqs. 3-6 will only deviate significantly from the classical when coherences are significant in which case Eqs. \ref{mcmatrix} and \ref{ppfinal} also cease to provide an accurate description of experimental results. This is expected to happen e.g., at short probe delay times when coherent processes are significant, but is out of the scope of the current investigation.

Finally, to gain further insight into the nature of the PP spectrum we compare in Fig. \ref{theo-in} the theoretical absolute transmission spectrum of the pump-excited system, $T\pu(\omega)$, with the linear spectrum $T^0(\omega)$.
Fig. \ref{theo-in} shows that the pump-induced molecular excited-state population leads to a probe response with three resonances, two of which are close to the linear LP frequency, and one that is slightly red-shifted from the linear UP. Hence, the reason much larger nonlinear signals are observed for LP relative to UP is that the bare molecule transition $\omega_{1\rightarrow2}$ is near-resonant with $\omega_{\text{LP}}$, while $\omega_{\text{UP}}$ is far off-resonant with the former. Thus, mechanical anharmonicity, which in the delocalized basis leads to polariton $-$ polariton, dark-state $-$ dark-state, and polariton $-$ dark-state interactions \cite{agranovich_hybrid_2011, carusotto_quantum_2013}, has a much weaker effect on UP compared to LP. Given that mechanical anharmonicity also represents the tendency of bonds to break at high-energies, we may conjecture that UP states will be more immune to bond-dissociation relative to LP. Roughly, this may also be understood from the point of view that the nonlinear signals of UP are much weaker than the LP for the studied system, so the UP behaves more like a harmonic oscillator than the LP. Noting that $\omega_{1\rightarrow 2} < \omega_{0\rightarrow 1}$ is a generic property of molecular vibrations, we can conclude that the weaker anharmonic character of the UP quasiparticles relative to LP is likely a general property of vibrational polaritons.

\emph{Acknowledgments.\textendash{}}
J.Y.Z and R.F.R acknowledge support from NSF CAREER award CHE:1654732 and generous UCSD startup funds. B.X and W.X are supported by the AFOSR Young Investigator Program Award, FA9550-17-1-0094. B.X thanks the Roger Tsien Fellowship from the UCSD Department of Chemistry and Biochemistry. J.Y.Z acknowledges discussions with Andrea Cavalleri on the various sources of anharmonicities.
\bibliography{lib}

\newpage{}

~

\newpage{}

 \onecolumngrid

\begin{center}
\textbf{\Large{}Supporting Information for Theory for Nonlinear Spectroscopy of Vibrational Polaritons}
\par\end{center}{\Large \par}

\global\long\def\theequation{S\arabic{equation}}
 \setcounter{equation}{0}

\global\long\def\thefigure{S\arabic{figure}}
 \setcounter{figure}{0}

Raphael F. Ribeiro$^{1}$, Adam D. Dunkelberger$^{2}$, Bo Xiang$^{3}$, Wei Xiong$^{1,3}$, Blake S. Simpkins$^{2}$, Jeffrey C. Owrutsky$^{2}$, Joel Yuen-Zhou$^{1}$

\emph{$^{1}$Department of Chemistry and Biochemistry, University
of California San Diego, La Jolla, CA}

\emph{raphaefr@gmail.com}

\emph{$^{2}$Chemistry Division, Naval Research Laboratory, Washington, District Of Columbia 20375}

\emph{$^{3}$Materials Science and Engineering Program, University of California, San Diego, La Jolla, CA 92093}

~
\date{\today}
\section{Linear absorption at strong-coupling}
In this section we derive the linear absorption spectrum of a system consisting of a microcavity strongly-coupled with a molecular vibration. This information is particularly significant for measurements that probe nonlinear transient polariton dynamics. For instance, in the main manuscript we assume that pumping the system leads to transient molecular excited-state population. This assumption is sufficient to obtain good agreement with polariton pump-probe experiments at sufficiently long probe delay time. This excited-state population might have arisen from polariton relaxation to dark-states and/or direct absorption of off-resonant photons by the system due to the finite cavity and molecular linewidths. While the detailed mechanism giving rise to the assumed transient population is outside the scope of the presented research, we dedicate this section to polariton absorption to shed light on one of the possible mechanisms for the generation of a pump-induced transient molecular excited-state population. 

The microcavity linear absorption spectrum $A(\omega)$ may be experimentally obtained from the relation $A(\omega) = 1 - T(\omega) - R(\omega)$, where $T(\omega)$ and $R(\omega)$ are the transmission and reflection spectra, respectively \cite{ebbesen_hybrid_2016}. The same approach can be carried theoretically given a method to compute $T(\omega)$ and $R(\omega)$. Within input-output theory \cite{gardiner_input_1985, gardiner1991quantum} $A(\omega)$ may also be obtained directly. This requires defining intracavity bath degrees of freedom (representing e.g., solvent or intramolecular modes) linearly-  and weakly-coupled to the molecular vibrations of interest; for simplicity, we also assume Markovian bath-molecule dynamics. In this case, the treatment of energy absorption by the molecular system is identical to that of transmission and reflection. In particular, the Hamiltonian for a model consisting of a linear medium inside a single-mode cavity may be written as ($\hbar=1$)

 \begin{align}  & H_0 = H_{\rm{c}} + H_{\text{m}}^{(0)} + H_{\rm{mc}} + H_{\text{lr}}+ H_{\text{clr}} + H_{\text{mb}}, \nonum 
 & H_{\text{c}} = \hbar \omega_c b^\dagger b,~~ H_{\text{m}}^{(0)} = \hbar \omega_0 \sum_{i=1}^N \amd \am, ~~ H_{\text{mc}} = \hbar g\sum_{i=1}^N (\amd b+b^\dagger \am), \nonum
   &H_{\text{lr}} = \int_{-\infty}^{\infty} \mathrm{d}\omega'\hbar \omega' \left[\left(b^\text{L}(\omega')\right)^\dagger b^\text{L}(\omega')+\left(b^\text{R}(\omega')\right)^\dagger b^\text{R}(\omega')\right], \nonum 
   &H_{\text{clr}} = \frac{\hbar\sqrt{\kappa/2}}{\sqrt{2\pi}}\int_{-\infty}^{\infty}\mathrm{d}\omega' \left[\left(b^\text{L}(\omega')\right)^\dagger b+\left(b^\text{R}(\omega')\right)^\dagger b + \hc \right], \nonum
   & H_{\text{mb}} = \sum_{i=1}^N\int_{-\infty}^{\infty} \mathrm{d}\omega' \hbar \omega' r_i^\dagger(\omega')r_i(\omega')+  \sum_{i=1}^N\frac{\hbar\sqrt{\gamma_m}}{\sqrt{2\pi}}\int_{-\infty}^{\infty} \mathrm{d}\omega' r_i^\dagger(\omega') a_i+ \hc,
  \end{align}
where $b$ is the cavity and $\am$ is the $i$th molecular annihilation operator, $\omega_c$ and $\omega_0$ are the cavity mode and molecular fundamental transition frequencies, $g$ is the single-state coupling between the cavity mode electric field and the molecular vibrations, $\kappa$ is the coupling between external and cavity electromagnetic field modes, $\{b^{\text{L(R)}}(\omega')\}$ are annihilation operators for the external left (right) electromagnetic field modes, and $r_i(\omega')$ is the bath mode of frequency $\omega'$ which is
weakly-coupled to molecule $i$. The modes $r_i(\omega')$ can be quickly integrated out by solving their Heisenberg equations of motion (eom) in terms of the molecular operators $a_i(t)$ \cite{gardiner1991quantum, ciuti_input-output_2006}. Assuming the molecular bath modes are in their vacuum state, the eom for the (normalized) average vibrational polarization $\braket{P(t)} = \frac{1}{\sqrt{N}}\sum_{i=1}^N \braket{a_i}$ is given by

\begin{align} &\frac{\mathrm{d}\braket{P(t)}}{\mathrm{d}t}  = -i \omega_0 \braket{P(t)} -  \frac{\gamma_m}{2}\braket{P(t)} - i g \sqrt{N} \braket{b(t)}. \end{align}
In the frequency domain, it follows that 
 \begin{align} \braket{P (\omega')} = \frac{-ig\sqrt{N}}{-i(\omega'-\omega_0)+\gamma_m/2}\braket{b(\omega')}.\end{align}
The molecular output modes $a_i^{\text{out}}(t)$ are defined as usual in terms of the future behavior of the vibrational bath operators
\cite{gardiner1991quantum,ciuti_input-output_2006}
\begin{equation} a_i^{\text{out}}(t) = \frac{i}{\sqrt{2\pi}}\int_{-\infty}^{\infty} \mathrm{d}\omega'  r_i(\omega',t_1) e^{-i \omega'(t-t_1)}, ~~t_1 > t, \end{equation}
where $r_i(\omega',t_1)$ denotes the bath operator $r_i$ of frequency $\omega'$ at a future time $t_1 > t$. It follows the input-output relation \cite{steck2007quantum, gardiner_input_1985} corresponding to each molecule can be written as
\begin{align} a_i^{\text{out}}(t) = \sqrt{\gamma_m} a_i(t) \implies
 P^{\text{out}}(t) = \sqrt{\gamma_m} P(t),\end{align}
 where in the last expression we used the definition of the molecular (linear) polarization output operator 
 \ben P^{\text{out}}(t) = \frac{1}{\sqrt{N}} \sum_{i=1}^N a_i^{\text{out}}(t). \een Given that the molecular vibrations provide the dominant source of cavity absorption, the steady-state absorption spectrum of the system can be well-approximated by
 \begin{equation}
 A(\omega) = \bigg|\frac{\braket{P^{\text{out}}(\omega)}}{\braket{b_{\text{in}}^L(\omega)}}\bigg|^2,
  \end{equation}
 where we assumed the external source acts on the system from the left.
 This can be computed  given a solution to the eom for the cavity mode annihilation operator $b(t)$ in terms of the input field $b_{\text{in}}^L(t)$. For the model discussed here, it follows that \cite{gardiner1991quantum,steck2007quantum}
 
\begin{align} \braket{b(\omega')} = \frac{\sqrt{\kappa/2}[i(\omega'-\omega_0)-\gamma_m/2] }{\left[i(\omega'-\omega_c)-\kappa/2\right]\left[i(\omega'-\omega_0)-\gamma_m/2\right] +Ng^2}\braket{b_{\in}^{L}(\omega')}.
\end{align}
Thus, the absorption spectrum implied by our model is given by
\begin{equation}
A(\omega) = \frac{g^2 N \gamma_m \kappa/2}{\bigg|(\omega-\omega_0+i\gamma_m/2)(\omega-\omega_c+i\kappa/2)-N g^2\bigg|^2}.
\end{equation}
Note that this approach gives only a rough approximation to the absorption coefficient, as the cavity leakage rate $\kappa$ is assumed independent of frequency and in-plane wave vector. For realistic cavities there are other loss mechanisms which we have also neglected here \cite{kavokin2017microcavities}. Additionally, only a single cavity-mode is considered,  and the variation of the in-plane wave vector of the cavity with detuning $\omega_c - \omega_0$ is entirely neglected. Thus, the above results are expected to hold only for a small neighborhood of frequencies around the cavity frequency $\omega_c$. In any case, knowing that $A(\omega)$ has poles at the LP and UP (complex) frequencies we can rewrite $A(\omega)$ in terms of the latter

\begin{equation} A(\omega) = \frac{N g^2 \gamma \kappa/2}{\left[(\omega-\omega\LP)^2+\gamma\LP^2/4\right]\left[(\omega-\omega\UP)^2+\gamma\UP^2/4\right]}, \end{equation}
so for the absorption at $\omega\LP$ and $\omega\UP$ we estimate
\begin{equation} A(\omega\LP) = \frac{2g^2 N \gamma_m \kappa}{\gamma\LP^2\left(\Omega_R^2 + \frac{\gamma\UP^2}{4}\right)},~~A(\omega\UP) = \frac{2g^2 N \gamma_m \kappa}{\gamma\UP^2\left(\Omega_R^2 + \frac{\gamma\LP^2}{4}\right)},\end{equation}
where $\Omega_R = \omega\UP - \omega\LP$. When $\omega_c - \omega_0 = 0$, $\Omega_R \approx 2g \sqrt{N}$. In the strong coupling limit it is common for $\gamma\LP/\Omega_R,\gamma\UP/\Omega_R  \rightarrow 0$. In this case, we can approximate
\begin{equation} A(\omega\LP) + A(\omega\UP) \approx \frac{2g^2N\gamma_m\kappa}{\Omega_R^2}\left(\frac{1}{\gamma\LP^2}+\frac{1}{\gamma\UP^2} \right) = \frac{g^2N\gamma_m\kappa}{2\sqrt{g^2 N + \frac{(\omega_c - \omega_0)^2}{4}}}\left(\frac{1}{\gamma\LP^2}+\frac{1}{\gamma\UP^2} \right). \end{equation}
Furthermore, when $(\gamma_m - \kappa)(\omega_c-\omega_0)/\Omega_R^2 \rightarrow 0$ it is also true that $\gamma\LP \approx \gamma\UP = (\gamma_m+\kappa)/2$. Thus, we see that  $A(\omega\LP) + A(\omega\UP)$ is maximized at zero detuning, i.e., when $\omega_c-\omega_0 = 0$. Another simple result following from the above considerations is the variation of the absorption at the dark-state frequency $\omega_0$ with respect to detuning:

\begin{equation} A(\omega_0) = \frac{g^2 N \gamma_m\kappa/2}{ N^2 g^4+\kappa^2 \gamma^2/16 +g^2N \kappa \gamma_m/2+\gamma_m^2(\omega_0-\omega_c)/4} \approx 
\frac{g^2 N \gamma_m\kappa/2}{ N^2 g^4+g^2N \kappa \gamma_m/2+\gamma_m^2(\omega_0-\omega_c)/4}. \label{dabs}\end{equation}
This indicates that $A(\omega_0)$ reaches a maximum when $\omega_0 \approx \omega_c$, though from Eq. \ref{dabs} its variation with detuning is slow.
 
\section{Interpretation of transient mode-coupling matrix}
We present in this section a detailed discussion of the mode-coupling matrix in Eq. 15 of the main text. First, recall that this matrix has eigenvalues that are determined by the poles of the probe output electric field
\begin{align} & \braket{b_{\out}^{\text{R}}}(\omega') = \frac{-i\frac{\kappa}{2} (\omega'-\omega_0 +i\gamma_1)(\omega'-(\omega_0-2\Delta)+i\gamma_3)}{(\omega'-\omega_c +i \kappa/2)(\omega'-\omega_0 +i\gamma_1)(\omega'-(\omega_0-2\Delta)+i\gamma_3)-N A(\omega',f\pu)}\braket{b_{\text{in}}^L}(\omega').\label{ppfinal}
\end{align}
As we will see, the elements of the mode-coupling matrix can be assigned a simpler interpretation when the molecular damping rate is assumed to be zero. Generally, the cavity leakage rate is much faster than the molecular relaxation, so the limit where the latter vanishes is reasonable when $f\pu$ (see main text, Eq. 7) is sufficiently small.

\subsection{Mode-coupling without vibrational relaxation}
If we take the molecular relaxation rate to be equal to zero  ($\gamma_m \rightarrow 0$), and $f\pu < 1/2$ (i.e., if the average population of molecules in the first excited-state is less than $50\%$ of the total number of molecules interacting with cavity photons), the following three-level system \textit{effective} coupling matrix has eigenvalues at the energies corresponding to the poles of the transient response given in Eq. \ref{ppfinal},

\begin{equation}
h(f\pu) =  \begin{pmatrix}
\omega_c-i\kappa/2 & g_1 \sqrt{N}\sqrt{1-2f\pu} & (g_1+g_3)\sqrt{2f\pu N} \\
g_1 \sqrt{N}\sqrt{1-2f\pu} & \omega_0 & 0 \\
(g_1+g_3)\sqrt{2f\pu N} & 0 & \omega_0 - 2\Delta
 \end{pmatrix}. \label{mcma}
\end{equation}
In order to understand the above, we look first at some limits: if $f\pu =0$, then 
\begin{equation}
h(f\pu)\bigg|_{f\pu=0} =  \begin{pmatrix}
\omega_c-i\kappa/2 & g_1 \sqrt{N} & 0 \\
g_1 \sqrt{N} & \omega_0 & 0 \\
0 & 0 & \omega_0 - 2\Delta
 \end{pmatrix}.
\end{equation}
The eigenvalues for this case are the frequencies of the (linear) LP, UP, and the bare molecule frequency $\omega_0 - 2\Delta$. There is no coupling between the cavity polarization and the $1\rightarrow 2$ matter transition, as expected.
If $\Delta = 0$, only electrical anharmonicity is non-vanishing and the mode-coupling matrix becomes
\begin{equation}
h(f\pu)\bigg|_{\Delta =0} =  \begin{pmatrix}
\omega_c-i\kappa/2 & g_1 \sqrt{N}\sqrt{1-2f\pu} & (g_1+g_3)\sqrt{2f\pu N} \\
g_1 \sqrt{N}\sqrt{1-2f\pu} & \omega_0 & 0 \\
(g_1+g_3) \sqrt{2f\pu N}& 0 & \omega_0 
 \end{pmatrix}.
\end{equation}
In this case, given the resonance between the $\omega_{10}$ and $\omega_{21}$ transitions, there exists a \textit{bright} and a \textit{dark} molecular state. The latter has frequency $\omega_0$, contains no photons, and hence does not contribute to the optical signal. Thus, we find that if only electrical anharmonicity exists, the linear response of the pumped system has two shifted polariton peaks (relative to the case with $f\pu=0$) (as verified explicitly in the l.h.s of Fig. 2 of our letter). If $g_3 = 0$, the system only has mechanical anharmonicity, whence it follows that
\begin{equation}
h(f\pu)\bigg|_{g_3 =0} =  \begin{pmatrix}
\omega_c-i\kappa/2 & g_1 \sqrt{N}\sqrt{1-2f\pu} & g_1\sqrt{2f\pu N} \\
g_1 \sqrt{N}\sqrt{1-2f\pu} & \omega_0 & 0 \\
g_1\sqrt{2f\pu N} & 0 & \omega_0 - 2\Delta
 \end{pmatrix}.
\end{equation}
The inclusion of mechanical anharmonicity leads to a system with three non-degenerate transient eigenvalues (resonances). As shown on the r.h.s of Fig. 2 of the main text, each eigenvector contains contributions from the photon part.

\subsubsection{Interpretation of couplings}
\def\rarrow{\rightarrow}
To interpret the off-diagonal elements corresponding to the coupling of cavity photons with molecular polarizations arising from the $0\leftrightarrow1$ and $1\rightarrow 2$ transitions, we first note that the presence of a transient population in the first molecular excited-state implies the molecular (linear) optical response $P(\omega)$ can be decomposed into two parts: 

\begin{equation} P(\omega) = P_{0\rarrow1}(\omega) + P_{1\rarrow2}(\omega), \end{equation}
where $P_{0\rarrow1}(\omega)$ and $P_{1\rarrow2}$ are the polarizations due to processes involving $0\leftrightarrow 1$, and 1$\rightarrow 2$ transitions, respectively. Additionally, $P_{0\rarrow1}(\omega)$ can be further decomposed into a sum of contributions,

\begin{equation} P_{0\rarrow1}(\omega) = P_{\text{NP}}(\omega) + P_{\text{GSB}}(\omega) + P_{\text{SE}}(\omega), \end{equation} 
where $P_{\text{NP}}(\omega)$ is the linear response of the molecular system when there is no transient population, $P_{\text{SE}}(\omega)$ accounts for stimulated emission processes, and $P_{\text{GSB}}(\omega)$ corresponds to the ground-state bleach contribution (see Fig. 1 of SI):
\begin{align}
& P_{\text{NP}}(\omega) = -i\frac{g_1^2N}{\omega-\omega_{01}}E(\omega), \\ & P_{\text{GSB}}(\omega)=  -i \frac{g_1^2N(-f\pu)}{\omega-\omega_{01}}E(\omega), \nonumber \\ & P_{\text{SE}}(\omega) = -i\frac{g_1^2N(-f\pu)}{\omega-\omega_{01}}E(\omega), \nonumber \\
 & P_{0\rarrow1}(\omega) = -i \frac{g_1^2N(1-2f\pu)}{\omega-\omega_{01}}E(\omega).
\end{align}
$P_{1\rarrow 2}(\omega)$ is the polarization component due to excited-state absorption of the transient population:
\begin{equation} P_{1\rarrow2}(\omega) = P_{\text{ESA}}(\omega) = -i \frac{2Nf\pu(g_1+g_3)^2}{\omega-\omega_{12}} E(\omega), \end{equation}
where we used  $\mu_{0\rarrow1} = g_1$ and $\mu_{1\rarrow2}=\sqrt{2}(g_1+g_3)$. 
\begin{figure}[b]
\includegraphics[scale=0.45]{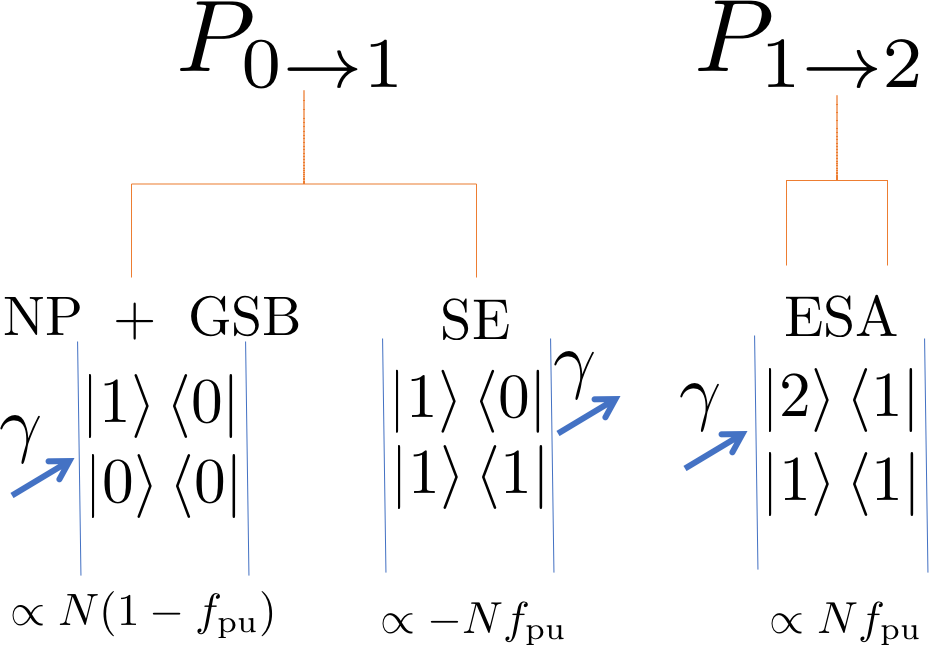}
\caption{Feynman diagrams for the processes occurring with a transient molecular system with $(1-f\pu)N$ molecules in the vibrational ground-state and $f\pu N$ molecules in the first excited-state. From left to right we show the contribution coming from ground-state molecules $P_{\text{NP}}(\omega)+P_{\text{GSB}}(\omega)$, the stimulated emission of excited-state molecules $P_{\text{SE}}(\omega)$, and the polarization due to excited-state absorption $P_{\text{ESA}}(\omega)$, respectively.}
\label{pol}
\end{figure}

Thus, we see that the molecular polarization has two linearly independent bare components $P_{0\rarrow1}(\omega)$ and $P_{1\rarrow2}(\omega)$. Each of these interact with cavity photons. The interaction strength is determined by the square root of  the \textit{numerator} of the response functions $P_{0\rarrow1}(\omega)$  and $P_{1\rarrow2}(\omega)$ obtained above. Thus, $P_{0\rarrow1}(\omega)$ interacts with an electric field of frequency $\omega$ with coupling constant $g_1 \sqrt{N(1-2f\pu)}$ and $P_{1\rarrow2}(\omega)$ interacts with the same field with coupling constant $(g_1+g_3) \sqrt{2Nf\pu}$. These are exactly the non-vanishing off-diagonal couplings of  $h(f\pu)$, which thus represent the interactions between cavity photons and the transient molecular system.\\

Now we understand \textit{why} when $f\pu = 1/2$ the $0\rightarrow 1$ polarization does not interact with light: $P_{0\rightarrow1}$ is the sum of the polarization induced in the system in the absence of any pumping $P_{\text{NP}}$, and the polarization components due to ground-state bleach and stimulated emission, which satisfy $P_{\text{GSB}} + P_{\text{SE}} =  -2f\pu P_{\text{NP}}$. Therefore, when $f\pu  = 50\%$, $P_{\text{GSB}} + P_{\text{SE}} =  -P_{\text{NP}}$ and $P_{0\rightarrow1}$ vanishes. In more intuitive terms, if $50\%$ of the molecules are in the first excited-state, the stimulated emission from the first excited-state coherently cancels the amplitude associated to ground-state absorption, so the total contribution of the $0\rightarrow 1$ transition to the matter polarization is zero. In this case, only the $1\rarrow 2$ effective dipole $P_{1\rarrow2}(\omega)$ contributes to the observed molecular polarization, and this is what the effective Hamiltonian in Eq. \ref{mcma} indicates. However, note that this conclusion is only true when the molecular excitations are not damped (see Sec.II.B)

\subsubsection{Case where $f\pu \geq 0.5$}

In this subsection we take a closer look at what happens with the mode-couplings in Eq. \ref{mcma} when $f\pu > 0.5$. In this case, the eigenvalues of $h(f\pu)$ can be analytically-continued when $f\pu > 1/2$, but the off-diagonal elements $[h(f\pu)]_{12} = [h(f\pu)]_{21} $ become purely imaginary.\\

To understand the unusual non-Hermitian structure of $h(f\pu)$ when $f\pu \geq 0.5$, note first that for small values of $f\pu$, all three eigenvalues of $h(f\pu)$ are complex and have a small \textit{negative} imaginary part. This agrees with the fact that the the cavity mode is lossy and the eigenmodes of $h(f\pu)$ inherit this property.

For $f\pu =0.5$ the mode-coupling matrix is given by
\begin{equation}
h(f\pu)\bigg|_{f\pu = 0.5} =  \begin{pmatrix}
\omega_c-i\kappa/2 & 0 & \sqrt{N}(g_1+g_3) \\
0 & \omega_0 & 0 \\
\sqrt{N}(g_1+g_3) & 0 & \omega_0 - 2\Delta
 \end{pmatrix},
\end{equation}
which implies that at $f\pu=0.5$, the $0\leftrightarrow 1$ molecular transitions become decoupled from light. Hence, $P_{0\rarrow1}$  is decoupled from the cavity photon, and 
$h(f\pu)$ has two complex eigenvalues with negative imaginary parts, and a single $\textit{real}$ eigenvalue $\omega_{01} = \omega_0$. Given that we expect a continuous behavior for the real and imaginary parts of the eigenvalues as a function of the matrix elements except at isolated singular points (where the couplings are non-analytic, or the eigenvalues cross), and all couplings are analytic except at $f\pu = 0$ and $f\pu = 1/2$ (no level crossing happens, either), it is reasonable to think that by switching $f\pu$ from small positive values to 0.5, the imaginary part of the eigenvalue of $h(f\pu)$ which approaches $\omega_0$ when $f\pu \rightarrow 0$ does so in a continuous way: it becomes \textit{less} negative as $f\pu \rightarrow 1/2$, and zero at $f\pu = 1/2$. When $f\pu > 0.5$ the mode with real frequency at $f\pu = 0$ acquires a positive imaginary part, which is a characteristic of a gain mode.

In the end, the purely imaginary coupling of $P_{0\rightarrow1}$ to the cavity photon is explained by the fact that we have jumbled absorption and emission of energy with frequency $\omega_{0}$ into a single polarization component $P_{0\rightarrow1}(\omega)$. Thus, our coarse-graining of the $P_{0\rightarrow1}(\omega)$ polarization is the culprit for non-hermiticity (at $f\pu > 1/2$) even when the cavity is lossless. An evidence for this hypothesis is that no unusual behavior happens for the cavity-$P_{1\rarrow2}(\omega)$ couplings ($P_{1\rarrow2}(\omega)$ only has a single-component due to excited-state absorption of $Nf\pu$, so according to our hypothesis its coupling with the cavity photon should be well-behaved for all values of $f\pu$ in the interval [0,1]).

\subsection{Mode-coupling with vibrational state relaxation, but no electrical anharmonicity ($g_3=0$)}

The effective mode-coupling matrix accounting for both bare molecule and cavity relaxation (and for simplicity assuming no electrical anharmonicity) is given by 
\begin{equation}
h(f\pu) =  \begin{pmatrix}
\omega_c-i\kappa/2 & g_1 \sqrt{N}\sqrt{1-2f\pu \frac{\kappa-\gamma_m}{\kappa-3\gamma_m}-u^2}-i u &-i\sqrt{2f\pu N}g_1\\
g_1 \sqrt{N}\sqrt{1-2f\pu \frac{\kappa-\gamma_m}{\kappa-3\gamma_m}-u^2}+i u & \omega_0 -i\gamma_m/2& 2\sqrt{2f\pu N}g_1\sqrt{\frac{\gamma_m}{\kappa-3\gamma_m}} \\
i\sqrt{2f\pu N}g_1 & 2\sqrt{2f\pu N}g_1\sqrt{\frac{\gamma_m}{\kappa-3\gamma_m}} & \omega_0 - 2\Delta -3i\gamma_m/2
 \end{pmatrix}, \kappa \neq 3\gamma_m,\label{mcmb}
\end{equation}
where $u = \sqrt{\frac{\gamma_m}{-6\gamma_m+2\kappa}} \left[\omega_c-(\omega_0-2\Delta)\right].$ The above is valid when  $\kappa \neq 3\gamma_m$, for the contrary implies the imaginary parts of $h_{11}$ and $h_{33}$ become degenerate; it leads to singular off-diagonal couplings in Eq. \ref{mcmb}, and thus invalidates the above matrix.

Eq. \ref{mcmb} is much more complicated than Eq. \ref{mcma}. Nonetheless, it can be easily verified that in the limit where $\gamma_m = 0$ the latter can be recovered from the former. Furthermore, as mentioned above, $f\pu = 1/2$ ceases to be a singular point of the off-diagonal couplings when $\gamma_m \neq 0$. Finally, we note that the molecular damping induces an effective interaction between the $P_{0\rightarrow1}$ and $P_{1\rightarrow 2}$ polarizations ($h_{23}$ and $h_{32}$ elements) which does not exist when $\gamma_m = 0$. A detailed study of the more complex features of Eq. \ref{mcmb} is beyond the scope of this work.

\end{document}